\documentclass[twocolumn,showpacs,preprintnumbers,amsmath,amssymb]{revtex4}

\usepackage{graphicx}
\usepackage{dcolumn}
\usepackage{bm}

\begin{document}

\title{Spinor Bose-Einstein Condensates of Rotating Polar Molecules}
\author{Y. Deng and S. Yi}
\affiliation{State Key Laboratory of Theoretical Physics, Institute
of Theoretical Physics, Chinese Academy of Sciences, P.O. Box 2735,
Beijing 100190, China}
\date{\today}

\begin{abstract}
We propose a scheme to realize a pseudospin-$1/2$ model of the $^{1}\Sigma(v=0)$ bialkali polar molecules with the spin states corresponding to two sublevels of the first excited rotational level. We show that the effective dipole-dipole interaction between two spin-$1/2$ molecules couples the rotational and orbital angular momenta and is highly tunable via a microwave field. We also investigate the ground state properties of a spin-$1/2$ molecular condensate. A variety of nontrivial quantum phases, including the doubly-quantized vortex states, are discovered. Our scheme can also be used to create spin-$1$ model of polar molecules. Thus, we show that the ultracold gases of bialkali polar molecules provide a unique platform for studying the spinor condensates of rotating molecules.
\end{abstract}


\maketitle

{\em Introduction}.---Recent experimental realizations of ultracold polar molecules in rovibrational ground state~\cite{KRb-exp,LiCs-exp,RbCs-exp,RbCs-exp2,RbCs-exp3} offer remarkable new frontiers for many areas of science, such as precision measurement~\cite{pre-measu1,pre-measu2,pre-measu3}, quantum information~\cite{qu-info}, quantum computation~\cite{qucompu}, ultracold collisions~\cite{krb-coll,lics-coll}, cold controlled chemistry~\cite{krb-chem,Miranda11}, and quantum simulation~\cite{spin-model,delmer06,rev-baranov}. Particularly, from the condensed matter perspective, the large permanent electric dipole moment and the ability to control the hyperfine states within a single rovibrational level~\cite{krb-hyper,neyen} make ultracold polar molecules an ideal platform for investigating strongly correlated many-body physics~\cite{Wall10,Rey11,Gorshkov11}. So far, the dipolar spin-exchange interactions~\cite{Yan13} and the many-body dynamics~\cite{Hazzard14} have been experimentally observed in lattice-confined ultracold KRb gases.

For rotating molecules, the net dipole moment in the lab frame vanishes in the absence of a dc electric field. Hence, in most theoretical many-body studies, a strong dc electric field is assumed to align polar molecules. Consequently, the rotational degrees of freedom is frozen. Even though, there exist multicomponent models by utilizing different rotational states, the number of molecules in each rotational state is independently conserved by the interactions. On the contrary, the spin-exchange contact interaction in atomic spinor Bose-Einstein condensates (BECs) results in rich magnetic phenomena~\cite{atom-spinor1,atom-spinor2,atom-spinor3}. Of particular interest, the magnetic dipole-dipole interaction (DDI) gives rise to spontaneous spin textures in dipolar spinor BECs~\cite{atom-dipspin-theo1,atom-dipspin-theo2,atom-dipspin-exp1,atom-dipspin-exp2}.

In this Letter, we show that a bialkali polar molecule in the electronic and vibrational ground state can be modeled as a pseudospin-$1/2$ molecule with the spin states corresponding to two hyperfine sublevels of the first excited rotational level. Remarkably, the effective DDI between molecules contains a rotation-orbit coupling term that is capable of inducing spin mixing. Thus a BEC formed by these spin-$1/2$ molecules represents a spinor BEC, instead of a two-component one. We also study the ground state phases of the spin-$1/2$ molecular BEC and demonstrate that rotation-orbit-coupled DDI gives rise to the doubly quantized vortex phases. Although BECs of polar molecules are studied in Refs.~\cite{molcond-wang,molcond-stoof,molcond-wilson}, the DDI considered in these works do not contain term that exchanges spin and orbital angular momentum.

{\em Model}.---To be specific, we consider a gas of $^{7}$Li$^{133}$Cs molecules in $^{1}\Sigma(v=0)$ state subjected to a bias magnetic field ${\mathbf B}=B\hat{\mathbf z}$. Each molecule has three angular momentum degrees of freedom: the rotation angular momentum ${\mathbf N}$ and the nuclear spins ${\mathbf I}_{1}$ and ${\mathbf I}_{2}$~\cite{Hutson08,Hutson09,Aldegund09,SM}. Its internal states can be characterized in the uncoupled basis $|M_{1}M_{2}NM_{N}\rangle$, where $M_{N}$ and $M_{i}$ are, respectively, the projections of ${\mathbf N}$ and ${\mathbf I}_{i}$ along the quantization $z$ axis. The Hamiltonian describes the internal degrees of freedom includes rotational $\hat{H}_{\rm rot}$, hyperfine $\hat{H}_{\rm hf}$, and Zeeman $\hat{H}_{Z}$ terms. Among them, the rotational term, $\hat H_{\rm rot}=B_{v}{\mathbf N}^{2}$, defines the largest intrinsic energy scale as the rotational constant $B_{v}$ is of order GHz. Since the rotational spectrum is anharmonic, we may focus on the lowest two rotational levels with $N=0$ and $1$, which are split by a energy $2B_{v}$.

\begin{figure}[tbp]
\includegraphics[width=0.85\columnwidth]{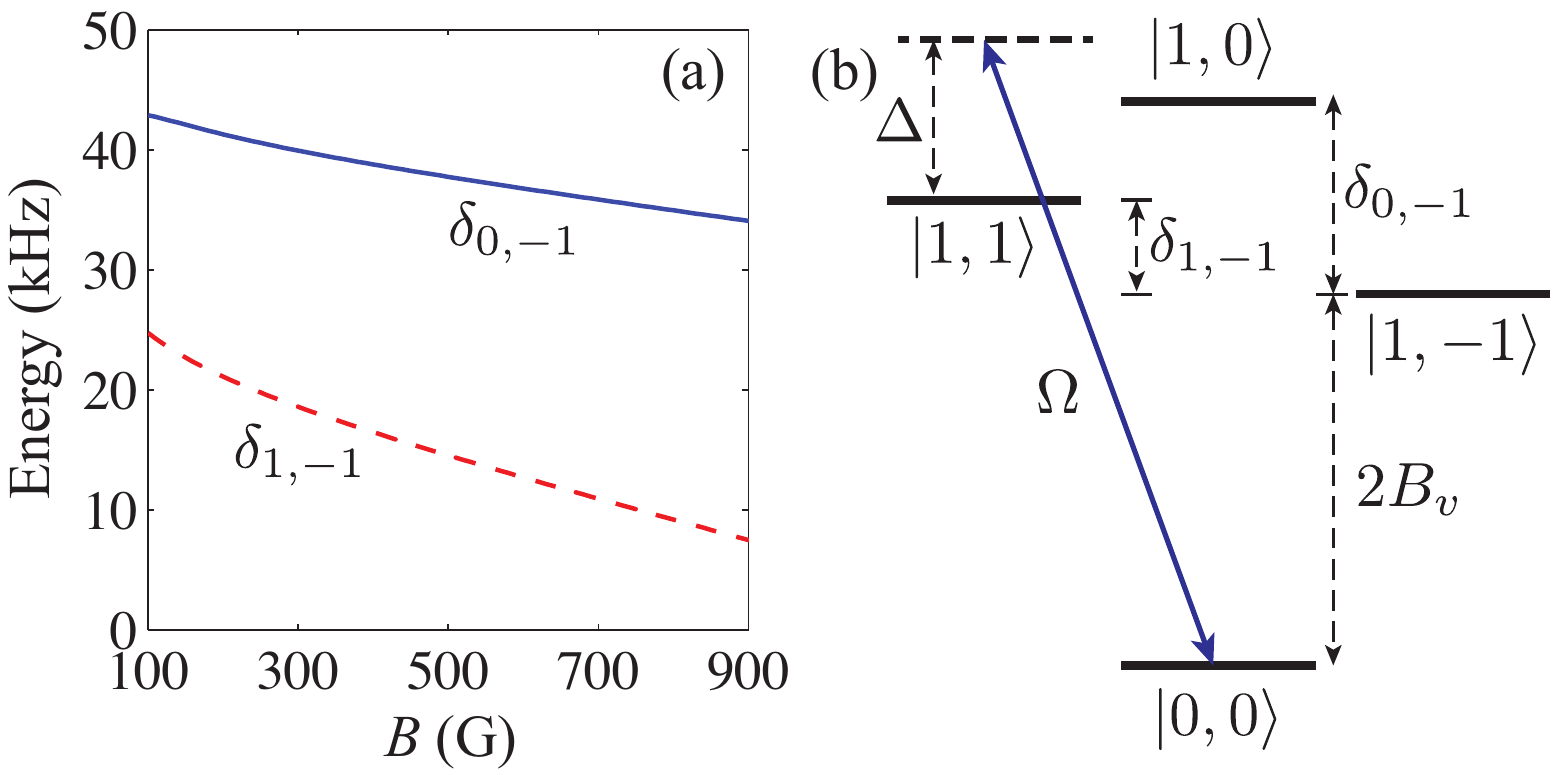}
\caption{(color online). (a) Hyperfine splittings as functions of the external magnetic field. (b) Level structure.}\label{scheme}
\end{figure}

Although the nuclear hyperfine interaction $\hat{H}_{\rm hf}$ mixes different internal states, it can be overcome by the Zeeman term $\hat H_{Z}$, which couples ${\mathbf B}$ to ${\mathbf N}$ and ${\mathbf I}_{i}$. For sufficiently strong magnetic field, the nuclear Zeeman effect dominates over the hyperfine interaction such that $M_{1}$ and $M_{2}$ become good quantum numbers. For LiCs, this magnetic field is around $40\,$G~\cite{Ran10}. Focusing on the lowest nuclear Zeeman levels ($M_{i}=I_{i}$) in the $N=0$ and $1$ manifolds, the relevant internal states reduces to $|N,M_{N}\rangle=|0,0\rangle$, $|1,0\rangle$, and $|1,\pm1\rangle$, which simplifies a rotating molecule to a four-level system. It can be verified that the hyperfine interaction is diagonal in this reduced four-level Hilbert space. Therefore, each of these four levels indeed possesses a definite quantum number $M_{N}$. In Fig.~\ref{scheme}(a), we plot the magnetic field dependence of the hyperfine splittings $\delta_{0,-1}=E_{|1,0\rangle}-E_{|1,-1\rangle}$ and $\delta_{1,-1}=E_{|1,1\rangle}-E_{|1,-1\rangle}$ for a LiCs molecule. As can be seen, the typical hyperfine splitting is around a few tens kHz for magnetic field in the range of $100$-$900\,$G. Figure~\ref{scheme}(b) shows the corresponding level structure.

To proceed further, we illuminate the molecules by a position-independent $\sigma^{+}$-polarized microwave field. The frequency of the microwave $\omega_{\rm mw}$ is assumed to be blue-detuned relative to the rotational splitting with a detuning $\Delta=2B_{v}/\hbar-\omega_{\rm mw}$, where the typical value of $\Delta$ is $100\,$MHz. The $|0,0\rangle\leftrightarrow|1,1\rangle$ transition is then induced by the microwave with Rabi frequency $\Omega$. Assuming that all molecules are initially prepared in the $|1,1\rangle$ state~\cite{krb-hyper}, level $|0,0\rangle$ can be adiabatically eliminated in the large detuning limit $|\Omega/\Delta|\ll1$. This procedure yields an effective level splitting $\delta=\delta_{1,-1}+\Omega^2/\Delta$ between the levels $|1,1\rangle$ and $|1,-1\rangle$. Moreover, under the condition $|\delta|\ll|\delta_{0,-1}|$, level $|1,0\rangle$ becomes well-separated from $|1,\pm1\rangle$, which eventually leads to the effective pseudospin-$1/2$ single-particle Hamiltonian in the rotating frame~\cite{SM}:
\begin{align}
{\hat { h}} = \frac{{\mathbf p}^2}{2m} {\hat {I}} {+} \hbar
\frac{\delta}{2}\hat{\sigma}_z, \label{single}
\end{align}
where $m$ is the mass of the molecule, ${\hat {I}}$ is the identity matrix, and for short-hand notation, we shall denote $|1,1\rangle$ and $|1,-1\rangle$ as $|\uparrow\rangle$ and $|\downarrow\rangle$, respectively. As analyzed below, $|\uparrow\rangle$ and $|\downarrow\rangle$ form a closed Hilbert space even in the presence of the molecule-molecule interactions. On a side note, when $\delta_{1,-1}\sim 0$ under an appropriate magnetic field, we may also realize a spin-$1$ model by coupling $|0,0\rangle$ and $|1,0\rangle$ states with a large detuned $\pi$-polarized microwave field.

In the second-quantized form, the single-particle Hamiltonian for the spin-$1/2$ molecules takes the form
\begin{eqnarray}
\hat {\cal H}_{0}=\sum_{\sigma}\int d{\mathbf r}\hat\psi^{\dag}_{\sigma}({\mathbf r})\left[\hat h_{\sigma\sigma}+U_{\rm opt}({\mathbf r})\right]\hat\psi_{\sigma}({\mathbf r}),
\end{eqnarray}
where $\hat\psi_{\sigma=\uparrow,\downarrow}$ is the field operator for the spin-$\sigma$ molecule and $U_{\rm opt}$ is the optical dipole trap, which is assumed to be spin-independent~\cite{neyen,Yan13}.

Next, the electric DDI between two molecules, described by the electric dipole moment operators $d\hat {\mathbf d}_{1}$ and $d\hat {\mathbf d}_{2}$, can be expressed as
\begin{align}
V_{\rm dd}({\mathbf R})=\frac{g_{d}}{|{\mathbf R}|^{3}}\left[\hat{\mathbf d}_{1}\cdot \hat{\mathbf d}_{2}-3(\hat{\mathbf d}_{1}\cdot\hat{\mathbf R})\,(\hat{\mathbf d}_{2}\cdot\hat{\mathbf R})\right],
\end{align}
where $g_{d}=d^{2}/(4\pi\epsilon_{0})$ is the DDI strength with $\epsilon_{0}$ being the electric permittivity of vacuum and $d$ the electric dipole moment ($5.5\,$Debye for LiCs), ${\mathbf R}$ is the vector connecting the molecules, and $\hat{\mathbf R}={\mathbf R}/|{\mathbf R}|$ is a unit vector. For a typical density $n=10^{13}\,{\rm cm}^{-3}$ of LiCs gas, the DDI energy, $g_{d}n$, is around $46\,$kHz, which justifies the elimination of the $|0,0\rangle$ level. Although there is no direct DDI between states in the $N=1$ manifold, effective DDI can be induced via the eliminated $|0,0\rangle$ state. As shown in the Supplemental material~\cite{SM}, in the rotating frame, the effective DDI that is time averaged over a period of $2\pi/\omega_{\rm mw}$ is
\begin{align}
&\hat{\cal H}_{\rm dd}=\hat {V}_{1}+\hat {V}_{2}+\hat V_{3},\\
&\hat {V}_{1}=\kappa g_d\sqrt{\frac{4\pi}{45}}\!\int\!
\frac{d {\bf r}_{1} d {\bf r}_{2}}{|{\mathbf R}|^{3}}Y_{20}(\hat{\mathbf R}):\!\hat{n}_{\uparrow}({\mathbf r}_{1})\hat{n}_{\uparrow}({\mathbf r}_{2})\!:,\nonumber\\
&\hat {V}_{2}=\kappa g_d\sqrt{\frac{4\pi}{45}}\!\int\!
\frac{d {\bf r}_{1} d {\bf r}_{2}}{|{\mathbf R}|^{3}}Y_{20}(\hat{\mathbf R}):\!\hat{S}_{+}({\mathbf r}_{1})\hat{S}_{-}({\mathbf r}_{2})\!:,\nonumber\\
&\hat {V}_{3}=\kappa g_d\sqrt{\frac{8\pi}{15}}\!\int\!
\frac{d {\bf r}_{1} d {\bf r}_{2}}{|{\mathbf R}|^{3}}\!\left[Y_{2,2}(\hat{\mathbf R}):\!\hat{n}_{\uparrow}({\mathbf r}_{1})\hat{S}_{-}({\mathbf r}_{2})\!:+{\rm h.c.}\right], \nonumber
\end{align}
where $\kappa=\Omega^{2}/\Delta^{2}$, $\hat n_{\sigma}=\hat\psi_{\sigma}^{\dag}\hat\psi_{\sigma}$, $\hat S_{-}=\hat\psi_{\downarrow}^{\dag}\hat\psi_{\uparrow}$, $\hat S_{+}=\hat S_{-}^{\dag}$, and $:\!\hat O\!:$ arranges operator in normal order. Clearly, $\hat{V}_{1}$ represents the density-density DDI between spin-$\uparrow$ molecules and $\hat{V}_{2}$ is the dipolar spin-exchange interaction between spin-$\uparrow$ and -$\downarrow$ molecules. Of particular interest, the dipolar density-spin interaction $\hat V_{3}$ couples the rotational and orbital angular momenta while keeps the total angular momentum conserved. 

Compared to the DDI appeared in other spin-$1/2$ models of polar molecules, selecting $|1,1\rangle$ and $|1,-1\rangle$ states gives rise to the rotation-orbit coupling term $\hat V_{3}$. In addition, the elimination of $|0,0\rangle$ state with a large detuned microwave field introduces a control knob $\kappa$ for the DDI. Throughout this work, we assume that $\kappa\leq 6\times10^{-4}$ to maintain a stable BEC. Consequently, the typical DDI energy between states in the $N=1$ manifold is around $\kappa g_{d}n\lesssim27\,$Hz, which further validates the assumption that $|1,0\rangle$ is well-separated from $|1,\pm1\rangle$. 

For completeness, we also include the collisional interaction term
\begin{eqnarray}
\hat {\cal H}_{\rm con}=\sum_{\sigma\sigma'}\frac{2\pi\hbar^{2} a_{\sigma\sigma'}}{m} \int
d{\mathbf r}\hat\psi_{\sigma}^{\dag}({\mathbf r})
\hat\psi_{\sigma'}^{\dag}({\mathbf r})\hat\psi_{\sigma'}({\mathbf
r}) \hat\psi_{\sigma}({\mathbf r}),\label{hcol}
\end{eqnarray}
where $a_{\sigma\sigma'}$ are the $s$-wave scattering lengths between the spin-$\sigma$ and -$\sigma'$ molecules. So far, the $s$-wave scattering lengths for LiCs molecules are unknown. For simplicity, we take the typical values of $a_{\uparrow\uparrow}=a_{\downarrow\downarrow}=a_{\uparrow\downarrow}=100a_{B}$ with $a_{B}$ being the Bohr radius. It can be estimated that the contact interaction energy is also of a few tens Hz. We remark that the spin structures presented below should not depend on the specific choice of $a_{\sigma\sigma'}$ as $\hat{\cal H}_{\rm con}$ conserves the number of molecules in individual spin state.

\begin{figure}[tbp]
\includegraphics[width=0.95\columnwidth]{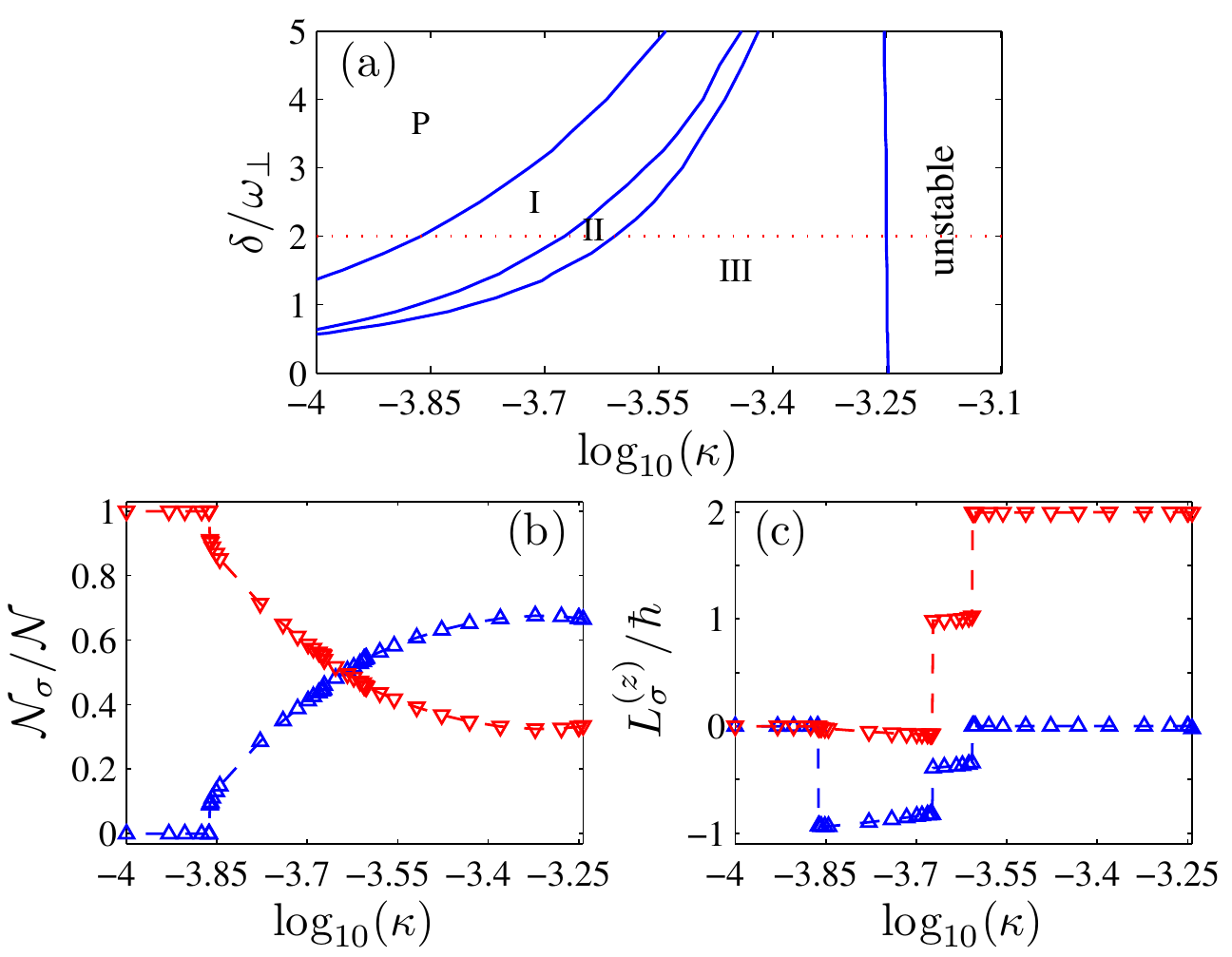}
\caption{(color online). (a) Phase diagram on the $\kappa$-$\delta$ parameter plane. (b) and (c) show, respectively, the molecule number and mean orbital angular momentum as functions of $\kappa$ for spin-$\uparrow$ ($\bigtriangleup$) and -$\downarrow$ ($\bigtriangledown$) states with $\delta=2\omega_{\perp}$. }
\label{fphase}
\end{figure}

{\em Quantum phases}.---We now turn to explore the ground state properties of a molecular BEC by using the mean-field theory. To this end, we replace the field operators $\hat\psi_{\sigma}$ by the condensate wave functions $\psi_{\sigma}\equiv\langle\hat{\psi}_{\sigma}\rangle$, which can be obtained by numerically minimizing the energy functional ${\cal F}[\psi_{\sigma},\psi_{\sigma}^{*}]=\langle \hat{\cal H}_{0}+\hat{\cal H}_{\rm dd}+\hat{\cal H}_{\rm con}\rangle$. More specifically, we consider a condensate of ${\cal N}=3.2\times10^5$ LiCs molecules trapped in a harmonic potential $U_{\rm opt}=m\omega_{\perp}^{2}(x^{2}+y^{2}+\gamma^{2}z^{2})/2$ with $\omega_{\perp}=(2\pi)10\,$Hz being the radial trap frequency and $\gamma=6.3$ being the trap aspect ratio. For simplicity, the condensate wave functions are decomposed into $\psi_{\sigma} ({\bf r}) = \phi_{\sigma}(x,y)\phi_z(z)$ with $\phi_z(z) = (\gamma/\pi \ell_{\perp}^2)^{1/4} e^{-\gamma z^2/2 \ell_{\perp}^2}$ and $\ell_{\perp}=\sqrt{\hbar/(m\omega_{\perp})}$. After integrating out the $z$ variable, the system simplifies to a quasi-two-dimensional one. Limited by validity of the spin-$1/2$ model, the numerical results presented below cover the parameter space $-10\leq \delta/\omega_{\perp}<10$ and $10^{-4}\leq\kappa\leq 6\times 10^{-4}$. 

Figure~\ref{fphase}(a) summarizes the phase diagram in the $\kappa$-$\delta$ parameter plane for a molecular BEC. The region denoted by P is the polarized phase and those labeled by I, II, and III represent three vortex phases. For $\kappa>5.7\times 10^{-4}$, the condensate becomes unstable. In Fig.~\ref{fphase}(b) and (c), we plot, for a fixed effective detuning $\delta=2\omega_{\perp}$, the $\kappa$ dependence of the molecule number ${\cal N}_{\sigma}=\int dxdy|\phi_{\sigma}|^{2}$  and the average orbital angular momentum $L^{(z)}_{\sigma}=-i\hbar{\cal N}_{\sigma}^{-1}\int dxdy\psi_{\sigma}^{*}\left(x\partial_{y}-y\partial_{x}\right)\psi_{\sigma}$ for each spin state. As can be seen, in the P phase, the spin-$\downarrow$ state is dominantly populated and the wave functions of both spin states are structureless; while in the vortex phases, molecules in either one or both spin states carry orbital angular momentum. Among the vortex phases, although ${\cal N}_{\sigma}$ vary smoothly with $\kappa$, the phase boundaries are clearly marked by $L_{\sigma}^{(z)}$. It should be noted that for large negative $\delta$, we still find that $L_{\downarrow}^{(z)}=2\hbar$ even though only spin-$\uparrow$ state is dominantly populated.

\begin{figure}[tbp]
\includegraphics[width=0.9\columnwidth]{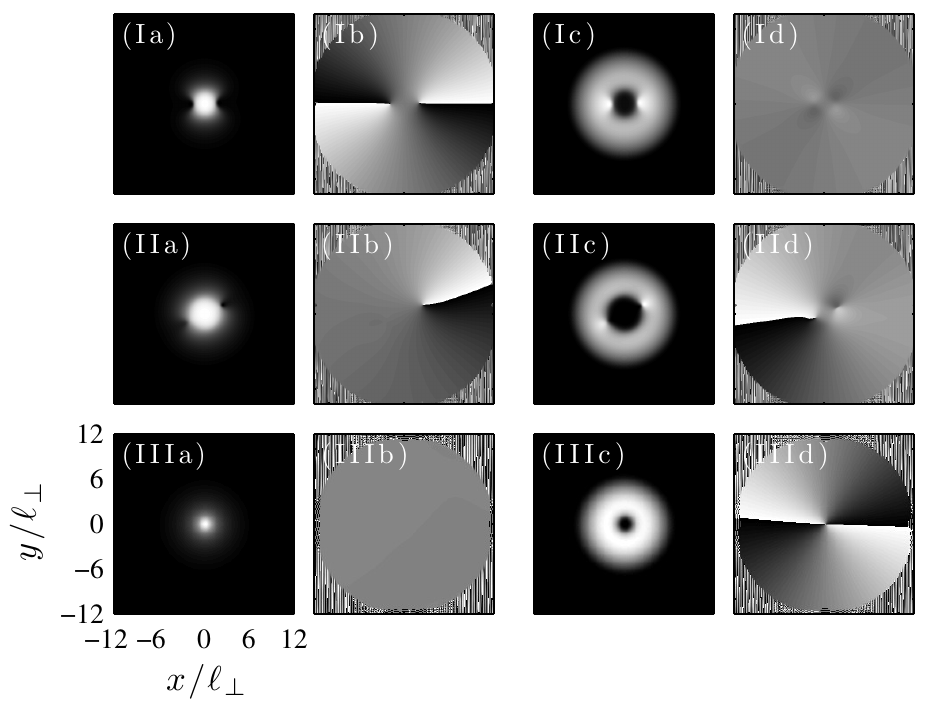}
\caption{Typical condensate wave functions for the vortex phases I (row 1), II (row 2), and III (row 3), corresponding to $\kappa=1.56\times 10^{-4}$, $2.3\times 10^{-4}$, and $4.94\times 10^{-4}$, respectively. The effective detuning used here is $\delta=2\omega_{\perp}$. Column 1 and 3 show the densities of the spin-$\uparrow$ and -$\downarrow$ molecules, respectively; column 2 and 4 show the corresponding phases.}\label{fwave}
\end{figure}

To gain more insight into the vortex phases, we present the wave functions for the phases I, II, and III in Fig.~\ref{fwave}. As shown in the phase plots, to lower the kinetic energy, only the less populated state is a vortex state when there is a large population imbalance. However, vortices appear in both components if ${\cal N}_{\uparrow}$ and ${\cal N}_{\downarrow}$ become comparable. The presence of the vortices can be understood from $\hat V_{3}$ in the DDI. By annihilating a spin-$\uparrow$ molecule and creating a spin-$\downarrow$ molecule, the rotational angular momentum is decreased by $2\hbar$. To ensure the total angular momentum conservation, the orbital angular momentum of spin-$\uparrow$ molecules must be larger than that of spin-$\downarrow$ molecules by $2\hbar$. Particularly, when one of the state is free of vortex (phases I and III), the other state must be a doubly quantized vortex state, in striking difference to the vortices in dipolar spin-1 atomic condensates~\cite{atom-dipspin-theo1}. 

As to the density profiles, spin-$\uparrow$ molecules always occupy the center of the trap with spin-$\downarrow$ molecules being pushed to the periphery. This configuration holds even if the scattering lengths are slightly tuned such that the contact interactions favor a miscible gas. In fact, aggregating at the trap center allows the spin-$\uparrow$ gas to stretch along the $z$ axis to the maximum extent such that the intraspecies DDI, $\hat V_{1}$, is lowered. The immiscibility, on the other hand, is induced by $\hat V_{2}$, as a miscible mixture in a pancake-shaped trap normally gives rise to a positive dipolar spin-exchange interaction energy due to the anisotropic nature of $Y_{2,0}$. Another consequence of this configuration is that the core of the vortex in the spin-$\uparrow$ molecules becomes off-axis such that the average orbital angular momentum $|L_{\uparrow}^{(z)}|/\hbar$ is less than $2$ and $1$ in phases I and II, respectively.

\begin{figure}[tbp]
\includegraphics[width=0.95\columnwidth]{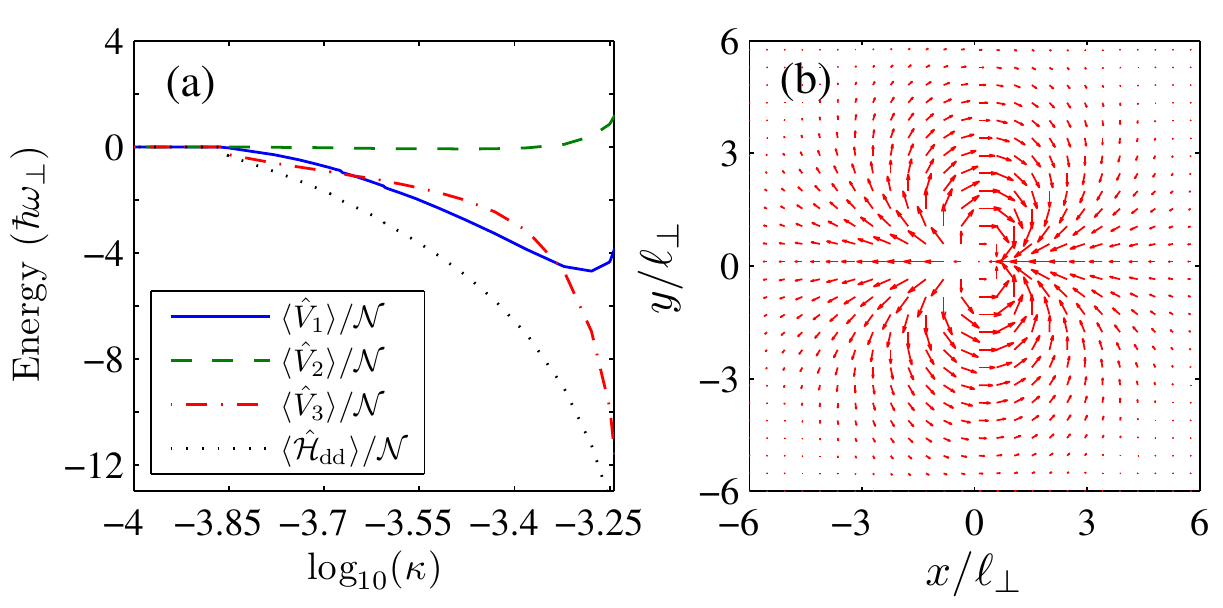}
\caption{(color online). (a) $\kappa$ dependence of the DDI energies per molecules for $\delta=2\omega_{\perp}$. (b) Typical planar spin structure for the vortex phases.}\label{fenergy}
\end{figure}

In Fig.~\ref{fenergy}(a), we plot the DDI energies as functions of $\kappa$. The negativity of $\langle\hat V_{1}\rangle$ indicates that spin-$\uparrow$ condensate is indeed of cigar shape whereas it is confined is a pancake-shaped trap. Moreover, the fact that $\langle\hat V_{2}\rangle$ roughly remains zero over a wide range of $\kappa$ is consistent with the immiscibility of the system. For $\langle\hat V_{3}\rangle$, it can be rewritten as
\begin{align}
\langle\hat V_{3}\rangle&=\kappa g_{d}\int\frac{d{\mathbf r}_{1}d{\mathbf r}_{2}}{R^{3}}n_{\uparrow}({\mathbf r}_{1})\sin^{2}\theta_{\mathbf R}\nonumber\\
&\qquad\times\left[s_{x}({\mathbf r}_{2})\cos(2\varphi_{\mathbf R})+s_{y}({\mathbf r}_{2})\sin(2\varphi_{\mathbf R})\right],
\end{align}
where $n_{\uparrow}=\langle \hat n_{\uparrow}\rangle$, $\theta_{\mathbf R}$ and $\varphi_{\mathbf R}$ are the polar and azimuthal angles of $\mathbf R$, respectively, $s_{x}=\frac{1}{2}(\langle\hat S_{+}\rangle+\langle\hat S_{-}\rangle)$, and $s_{y}=\frac{1}{2i}(\langle\hat S_{+}\rangle-\langle\hat S_{-}\rangle)$. Clearly, $\hat V_{3}$ should align the planar spin ${\mathbf s}_{\perp}=(s_{x},s_{y})$ such that $\langle \hat V_{3}\rangle$ is always negative. In fact, as shown in Fig.~\ref{fenergy}(b), ${\mathbf s}_{\perp}$ always forms a spin vortex with winding number $2$ in the vortex phases. In competing with $\hat V_{2}$, it is energetically favorable to have a large overlap between $\psi_{\uparrow}$ and $\psi_{\downarrow}$ for $\hat V_{3}$. Consequently, $\langle\hat V_{1}\rangle$ and $\langle\hat V_{2}\rangle$ significantly increases with $\kappa$ in the strong DDI regime. Since $\langle \hat V_{3}\rangle$ also depends on $n_{\uparrow}$, it explain why ${\cal N}_{\uparrow}$ continuously grows with $\kappa$ [Fig.~\ref{fphase}(b)], instead of being saturated at around ${\cal N}/2$. Finally, it is worthwhile to point out that the condensate becomes unstable when the DDI interaction energy is comparable to the contact interaction energy. The critical value of $\kappa$ is insensitive to $\delta$, as the the spins are free to rearrange themselves to minimize the dipolar interaction energy.

{\em Experimental feasibility}.---The realization of the proposed model requires the molecules to possess a large hyperfine splitting such that the effective DDI would not mix the unwanted rotational sublevel. In fact, the nuclear electric quadrupole coupling constants for all bialkali polar molecules with known molecular parameters are of order $100\,$kHz~\cite{Hutson08}, indicating that the proposed scheme is also applicable to other bialkali polar molecules. 

As to the experimental detection, similar to imaging an atomic spinor condensate, we may construct a Stern-Gerlach apparatus by utilizing the rotational Zeeman shift, $-g_r \mu_N {\mathbf N}\cdot {\mathbf B}$, where $g_{r}$ is the rotational $g$-factor of the molecule and $\mu_{N}$ is the nuclear magnetic moment. It can be estimated that for a modest magnetic field gradient of a few ${\rm T/m}$, the spin-$\uparrow$ and -$\downarrow$ states of the LiCs molecules are spatially separated after $200\,$ms free expansion and can be directly observed with absorption image measurement~\cite{imag-wang}.

{\em Conclusions}.---We have demonstrated that a rotating bialkali polar molecule can be modeled as a pseudospin-$1/2$ particle by utilizing dc electric and microwave fields. In this model, a control knob for the effective molecular DDI is naturally introduced, which can be used to stabilize the condensates of polar molecules with large electric dipole moment. More remarkably, the rotation-orbit coupling term in the effective DDI gives rise to doubly quantized vortex phases of the molecular condensate. Finally, the proposed scheme also works for the ultracold gases of fermionic polar molecules, in which the effective DDI may leads to exotic superfluid pairings.

This work was supported by the National 973 program (Grant No. 2012CB922104) and the NSFC (Grant Nos. 11025421, 11434011, and 11121403).

{\em Note added}.---During preparation of the manuscript, we becomes aware of the work by Wall {\it et al}.~\cite{wall14} for realizing unconventional quantum magnetism with symmetric top molecules, in which the effective DDI also exchanges the spin and orbital angular momentum.

\clearpage
\setcounter{equation}{0} 
\setcounter{figure}{0}
\setcounter{table}{0} 
\makeatletter

\begin{widetext}

\section*{Supplemental material for ``Spinor Bose-Einstein condensates of rotating polar molecules''}

\subsection{Hyperfine structure of $^{1}\Sigma$ bialikali molecules}
In order to simplify the internal states of a $^{1}\Sigma$ bialikali molecule to a four-level system, we have to consider the hyperfine structure of the molecule. When subjected to a bias magnetic field ${\mathbf B}=B\hat{\mathbf z}$, the Hamiltonian characterizing the internal degrees of freedom of a ${}^{1}\Sigma$ diatomic molecule is~\cite{book,SM-Hutson08,SM-Aldegund09}
\begin{align}
\hat{H_{\rm in}}= \hat{H}_{\rm rot} + \hat{H}_{\rm hf} + \hat{H}_Z,\label{smhin}
\end{align}
where the rotational term takes the form
\begin{align}
\hat{H}_{\rm rot}=B_{v}{\mathbf N}^2 \label{smhrot}
\end{align}
with $B_{v}$ being the rotational constant. Clearly, $\hat H_{\rm rot}$ diagonal in the basis $\{|M_{1}M_{2}NM_{N}\rangle\}$ as 
\begin{align}
\hat H_{\rm rot}|NM_{N}\rangle=B_{v}N(N+1)|NM_{N}\rangle.
\end{align}
For the reason stated in the main text, we may focus on the $N=0$ and $1$ rotational levels. This assumption is particularly valid in the absence of a dc electric field.

The nuclear hyperfine interaction contains four contributions: nuclear electric quadrupole interaction $\hat H_{Q}$, nuclear spin-rotation interaction $\hat H_{IN}$, tensor $\hat H_{t}$ and scalar $\hat H_{\rm sc}$ nuclear spin-spin interactions. Explicitly, the hyperfine Hamiltonian can be expressed as
\begin{align}
\hat{H}_{\rm hf}&=\hat H_{Q}+\hat H_{IN}+\hat H_{t}+\hat H_{\rm sc}\nonumber\\
&=\sum_{i=1}^{2}\frac{\sqrt{6}(eQ_{i}q_{i})}{4I_{i}(2I_{i}-1)}T^{(2)}({\mathbf C})\cdot T^{(2)}({\mathbf I}_{i},{\mathbf I}_{i})+\sum_{i=1}^2c_i{\mathbf N}\cdot {\mathbf I}_i -c_{3}\sqrt{6}\,T^{(2)}({\mathbf C})\cdot T^{(2)}({\mathbf I}_{1},{\mathbf I}_{2})+c_{4}{\mathbf I}_{1}\cdot{\mathbf I}_{2}, \label{smhhf}
\end{align}
where $T^{(2)}({\mathbf C})$ is the second order unnormalized spherical harmonic with components $T_{q}^{(2)}({\mathbf C})\equiv C_{q}^{(2)}(\theta,\varphi)=\sqrt{\frac{4\pi}{5}}Y_{2,q}(\theta,\varphi)$ with $(\theta,\varphi)$ being the spherical coordinate and $T^{(2)}({\mathbf I}_{i},{\mathbf I}_{j})$ represents the spherical tensor operator of rank $2$, formed by the vector operators ${\mathbf I}_{i}$ and ${\mathbf I}_{j}$. Moreover, $eQ_{i}$ is the electric quadrupole moment of nucleus $i$, $q_{i}$ characterizes the negative of the electric field gradient at nucleus $i$, $c_{i}$ represents the strength of the nuclear spin-rotation coupling for the $i$th nucleus, and $c_{3}$ and $c_{4}$ are, respectively, the strengths of the nuclear tensor and scalar spin-spin interaction. In Tab.~\ref{tabl}, we list the molecular parameters for several bialikali molecules. For convenience, we also present the matrix elements of the hyperfine interaction in the uncoupled basis~\cite{SM-Aldegund09,SM-Gorshkov11,SM-Wall10}:
\begin{align}
\langle M_{1}M_{2}NM_{N}|\hat H_{Q}|M_{1}'M_{2}'N'M_{N}'\rangle=&\sum_{i=1,2}\frac{(eQq)_{i}}{4}\delta_{M_{\bar i}M_{\bar i}'}\sum_{p}(-1)^{p-M_{N}+I_{i}-M_{i}}\sqrt{(2N+1)(2N'+1)}\nonumber\\
&\times\begin{pmatrix}N&2&N'\\-M_{N}&p&M_{N}'\end{pmatrix}\begin{pmatrix}I_{i}&2&I_{i}\\-M_{i}&-p&M_{i}'\end{pmatrix}\begin{pmatrix}N&2&N'\\0&0&0\end{pmatrix}\begin{pmatrix}I_{i}&2&I_{i}\\-I_{i}&0&I_{i}\end{pmatrix}^{-1},\label{smmathq}\\
\langle M_{1}M_{2}NM_{N}|\hat H_{IN}|M_{1}'M_{2}'N'M_{N}'\rangle=&\;\delta_{NN'}\sum_{q}(-1)^{q+N-M_{N}}\sqrt{N(N+1)(2N+1)}\begin{pmatrix}N&1&N\\-M_{N}&q&M_{N}'\end{pmatrix}\nonumber\\
&\times\sum_{i=1,2}c_{i}(-1)^{I_{i}-M_{i}}\delta_{M_{\bar i}M_{\bar i}'}\sqrt{I_{i}(I_{i}+1)(2I_{i}+1)}\begin{pmatrix}I_{i}&1&I_{i}\\-M_{i}&-q&M_{i}'\end{pmatrix},\label{smmathin}\\
\langle M_{1}M_{2}NM_{N}|\hat H_{t}|M_{1}'M_{2}'N'M_{N}'\rangle=&-c_{3}\sqrt{6}\sqrt{I_{1}(I_{1}+1)(2I_{1}+1)}\sqrt{I_{2}(I_{2}+1)(2I_{2}+1)}\sqrt{(2N+1)(2N'+1)}\nonumber\\
&\times\begin{pmatrix}N&2&N'\\0&0&0\end{pmatrix}\sum_{p}(-1)^{p-M_{N}+I_{1}-M_{1}+I_{2}-M_{2}}\begin{pmatrix}N&2&N'\\-M_{N}&p&M_{N}'\end{pmatrix}\nonumber\\
&\times\sum_{m}\langle 1,m;1,-p-m|2,-p\rangle\begin{pmatrix}I_{1}&1&I_{1}\\-M_{1}&m&M_{1}'\end{pmatrix}\begin{pmatrix}I_{2}&1&I_{2}\\-M_{2}&-p-m&M_{2}'\end{pmatrix},\label{smmatht}\\
\langle M_{1}M_{2}NM_{N}|\hat H_{\rm sc}|M_{1}'M_{2}'N'M_{N}'\rangle=&\;c_{4}\delta_{NN'}\delta_{M_{N}M_{N}'}\sqrt{I_{1}(I_{1}+1)(2I_{1}+1)}\sqrt{I_{2}(I_{2}+1)(2I_{2}+1)}\nonumber\\
&\times(-1)^{I_{1}-M_{1}+I_{2}-M_{2}}\sum_{p}(-1)^{p}\begin{pmatrix}I_{1}&1&I_{2}\\-M_{1}&p&M_{1}'\end{pmatrix}\begin{pmatrix}I_{2}&1&I_{2}\\-M_{2}&-p&M_{2}'\end{pmatrix}.\label{smmathsc}
\end{align}
where $\bar i=3-i$. One should note that (i) $\hat H_{IN}$ does not couple states with different $N$ and it plays a very small role in the spectra due to the smallness of the parameters $c_{1}$ and $c_{2}$; (ii) $\hat H_{t}$ often has a negligible effect as $c_{3}$ is usually of order $10$-$100\,$Hz; (iii) $\hat H_{Q}$ does not affect the $N=0$ level, however, it dominates for the $N=1$ level; (iv) $\hat H_{\rm sc}$ splits the various levels according to their total nuclear spin $I$ and it is the dominant hyperfine contribution for $N=0$ in the absence of external electric field.

Finally, the Hamiltonian describes the Zeeman term is
\begin{align}
\hat{H}_{Z}&=- g_r \mu_N {\mathbf N}\cdot {\mathbf B} -
\sum_{i=1}^2 g_i \mu_N {\mathbf I}_i\cdot {\mathbf
B}(1-\sigma_i)\nonumber .
\end{align}
where $\mu_N$ is the nuclear magnetic moment, $g_r$ is the rotational $g$-factor of the molecule, $g_i$ is the nuclear $g$-factor for the $i$th nucleus, and $\sigma_{i}$ is the nuclear shielding parameter. 
\begin{table}[h]
\tabcolsep 1pt \caption{Molecular parameters for bialkali polar molecules.  Subscripts 1 and 2 refer to the less electronegative atom and to the more electronegative one~\cite{SM-Hutson08,SM-Ran10,krb-polari}.} \vspace*{-12pt}
\begin{center}
\def\temptablewidth{0.7\textwidth}
{\rule{\temptablewidth}{1.5pt}}
\begin{tabular*}{\temptablewidth}{@{\extracolsep{\fill}}lcccc}
Molecule & $^7$Li$^{133}$Cs & $^{40}$K$^{87}$Rb & $^{41}$K$^{87}$Rb & $^{87}$Rb$^{133}$Cs\\
\hline
$I_{1}$             & $3/2$    & $4$       & $3/2$    & $3/2$\\
$I_{2}$             & $7/2$    & $3/2$     & $5/2$    & $7/2$\\
$g_{1}$             & $2.171$  & $-0.324$  & $0.143$  & $1.834$\\
$g_{2}$             & $0.738$  & $1.834$   & $0.541$  & $0.738$\\
$B_{v}$\;(GHz)      & $5.636$  & $1.114$   & $1.104$  & $0.504$\\
$(eqQ)_{1}$\;(kHz)  & $18.5$   & $452$     & $-298$   & $-872$\\
$(eqQ)_{2}$\;(kHz)  & $188$    & $-1308$   & $-1520$  & $51$ \\
$\sigma_{1}$\;(ppm) & $108.2$  & $1321$    & $1321$   & $3531$ \\
$\sigma_{2}$\;(ppm) & $6242.5$ & $3469$    & $3469$   & $6367$  \\
$c_1$\;(Hz)         & $32$     & $-24.1$   & $10.4$   & $98.4$\\
$c_{2}$\;(Hz)       & $3014$   & $420.1$   & $413.1$  & $194.1$\\
$c_3$\;(Hz)         & $140$    & $-48.2$   & $21.3$   & $192.4$\\
$c_4$\;(Hz)         & $1610$   & $-2030.4$ & $896.2$  & $17345.4$\\
$g_r$               & $0.0106$ & $0.0140$  & $0.0138$ & $0.0062$\\
$d$\;(Debye)        & $5.52$   & $0.566$    & $0.566$   & $1.25$
\end{tabular*}\label{tabl}
{\rule{\temptablewidth}{1pt}}
\end{center}
\end{table}

Once $M_{i}$ are fixed, it can be seen from Eq. (\ref{smmathq})-(\ref{smmathsc}) that the hyperfine interaction is diagonal in Hilbert space formed by $N=0$ and $1$ rotational states. Within this Hilbert space, the Hamiltonian describes the internal degrees of the molecule now becomes
\begin{align}
\hat H_{\rm in}=2B_{v}\sum_{q=0,\pm1}|1,q\rangle\langle 1,q|+\hbar\delta_{1,-1}|1,1\rangle\langle 1,1|+\hbar\delta_{0,-1}|1,0\rangle\langle 1,0|.
\end{align}

\subsection{Spin-$1/2$ single-particle Hamiltonian}
Here we show that by applying a $\sigma^{+}$-polarized microwave field, 
\begin{align}
{\mathbf E}(t) = E_{\rm wm}e^{-i\omega_{\rm mw} t}{\mathbf e}_{1}+ {\rm c.c},
\end{align}
the reduced four-level system can be further simplified to a psuodospin-$1/2$ one, where $E_{\rm wm}$ is the position-independent amplitude and the spherical vectors are defined as ${\mathbf e}_{0}=\hat{\mathbf z}$ and ${\mathbf e}_{\pm1}=\mp(\hat{\mathbf x}\pm i\hat{\mathbf y})/\sqrt{2}$ in the space-fixed frame, representing the $\sigma^{+}$ ($\hat{\mathbf e}_{1}$), $\pi$ ($\hat{\mathbf e}_{0}$), and $\sigma^{-}$ ($\hat{\mathbf e}_{-1}$) polarization of the microwave with respect to the quantization $z$ axis. The microwave field couples to the dipole moment operator,  ${\mathbf d}=d\hat{\mathbf d}$, of the molecule through the Hamiltonian 
\begin{align}
\hat H_{\rm mw}&=-{\mathbf d}\cdot {\mathbf E}(t)\nonumber\\
&=-E_{\rm mw}(d_{1}e^{-i\omega_{\rm mw}t}+d_{1}^{\dag}e^{i\omega_{\rm mw}t})
\end{align}
where $d_{q}={\mathbf d}\cdot{\mathbf e}_{q}=dC_{q}^{(1)}(\theta,\varphi)$ with $C_{q}^{(1)}(\theta,\varphi)=\sqrt{\frac{4\pi}{3}}Y_{1,q}(\theta,\varphi)$ being the permanent dipole moment of the molecule. Now, the single-molecule Hamiltonian becomes
\begin{align}
\hat H_{\rm in}+\hat H_{\rm mw}&=2B_{v}\sum_{q=0,\pm1}|1,q\rangle\langle 1,q|+\hbar\delta_{1,-1}|1,1\rangle\langle 1,1|+\hbar\delta_{0,-1}|1,0\rangle\langle 1,0|-\hbar\Omega\left(e^{-i\omega_{\rm mw}t}|1,1\rangle\langle 0,0|+{\rm h.c.}\right),\label{smhinmw}
\end{align}
where 
\begin{align}
\hbar\Omega=E_{\rm mw}\langle 1,1|d_{1}|0,0\rangle=\frac{dE_{\rm mw}}{\sqrt{3}}
\end{align}
is the Rabi frequency.

To proceed further, we rewrite the Hamiltonian (\ref{smhinmw}) in terms of the annihilation operators $\hat \psi_{NM_{N}}$ as
\begin{align}
\hat{H}_{\rm in}+\hat{H}_{\rm mw}=2B_{v}\sum_{q=0,\pm1}\hat{\psi}^\dag_{1q}\hat{\psi}_{1q}+ \hbar \delta_{1,-1} \hat{\psi}^\dag_{11}\hat{\psi}_{11}+  \hbar \delta_{0,-1}
\hat{\psi}^\dag_{10}\hat{\psi}_{10}-\hbar \Omega \left(\hat{\psi}^\dag_{11}\hat{\psi}_{00}e^{-i\omega_{\rm mw} t} +{\rm h.c.}\right).
\end{align}
We note that the spontaneous emissions of the excited rotational levels ($N=1$) are ignored due to the long lifetime of the rotational state. By introducing a rotating frame defined by the unitary transformation 
\begin{align}
{\cal U} = \exp(-i\hat{H}'t/\hbar)\label{smunitary}
\end{align}
with $\hat{H}' =\hbar\Delta
\hat{\psi}^\dag_{00}\hat{\psi}_{00}+2B_{v}\sum_{q=0,\pm1}
\hat{\psi}^\dag_{1q}\hat{\psi}_{1q}$, we obtain the time-independent Hamiltonian
\begin{align}
\hat H_{\rm in}+\hat{H}_{\rm mw} &\rightarrow {\cal U}^\dag(\hat{H}_{\rm in}+\hat H_{\rm mw}){\cal U} - i\hbar
{\cal U}^\dag\frac{\partial}{\partial t}{\cal U}, \nonumber \\
&=\hbar\left[-\Delta\hat{\psi}^\dag_{00}\hat{\psi}_{00}-\Omega\left(\hat{\psi}^\dag_{11}\hat{\psi}_{00}+
h.c.\right)+ \delta_{1,-1}\hat{\psi}^\dag_{11}\hat{\psi}_{11} +
\delta_{0,-1}\hat{\psi}^\dag_{10}\hat{\psi}_{10} \right].
\end{align}
In the rotating frame, the equations of motion for the annihilation operators are
\begin{align}
i\dot{\hat \psi}_{00} &=  -\Delta{\hat \psi}_{00} -\Omega{\hat
\psi}_{11}, \nonumber \\
i\dot{\hat \psi}_{11} &=  \delta_{1,-1}{\hat \psi}_{11} -\Omega{\hat \psi}_{00}, \nonumber \\
i\dot{\hat \psi}_{10} &=  \delta_{0,-1}{\hat \psi}_{10}, \nonumber \\
i\dot{\hat \psi}_{1-1} &= 0.\nonumber
\end{align}
Assuming that all molecules are initially prepared in the $|1,1\rangle$ state and $|\Delta|\gg|\Omega|,|\delta_{1,-1}|, |\delta_{0,-1}|$, the $|0,0\rangle$ level can be adiabatically eliminated to yield
\begin{align}
\hat\psi_{00}=-\frac{\Omega{\hat \psi}_{11}}{\Delta}.\label{smae}
\end{align}
The adiabatic elimination of the $|0,0\rangle$ level also induces a Stark shift, $\Omega^{2}/\Delta$, to the $|1,1\rangle$ level, such that the effective single-particle Hamiltonian becomes
\begin{align}
\hat H_{\rm in}+\hat H_{\rm mw}=\hbar\left(\delta\hat{\psi}^\dag_{11}\hat{\psi}_{11} +
\delta_{0,-1}\hat{\psi}^\dag_{10}\hat{\psi}_{10} \right)
\end{align}
with $\delta=\delta_{1,-1}+\Omega^2/\Delta$. Choosing $\delta/2$ as the origin of the energies, the above Hamiltonian can be rewritten as
\begin{align}
\hat H_{\rm in}+\hat H_{\rm mw}&=\hbar\left[\frac{\delta}{2}\hat{\psi}^\dag_{11}\hat{\psi}_{11} +
\left(\delta_{0,-1}-\frac{\delta}{2}\right)\hat{\psi}^\dag_{10}\hat{\psi}_{10}-\frac{\delta}{2}\hat{\psi}^\dag_{1-1}\hat{\psi}_{1-1} \right].\nonumber
\end{align}
As analyzed in the main text, by choosing an appropriate Stark shift or magnetic field strength, we may realize the condition, $|\delta|\ll|\delta_{0,-1}|$. Consequently, the $|1,0\rangle$ states becomes well-separated from the nearly degenerate $|1,\pm1\rangle$ states even in the presence of the molecule-molecule interactions (see below). After dropping the $|1,0\rangle$ state and taking into account the center of mass motion, we finally obtain the effective spin-$1/2$ single-molecule Hamiltonian [Eq.~(1) in the main text].

\subsection{Dipole-dipole interactions in the reduced spin-$1/2$ system}
For convenience, let us first write down the matrix elements of the dipole moment operator ${\mathbf d}$ in the rotational state basis $|NM_{N}\rangle$:
\begin{align}
\langle NM_N|d_q|N'M_N'\rangle &= (-1)^{2N-M_N}d\sqrt{(2N+1)(2N'+1)}\begin{pmatrix}
N & 1 & N' \\
-M_{N}  & q  & M_{N}' \end{pmatrix}
\begin{pmatrix}
N & 1 & N' \\
0  & 0 & 0
\end{pmatrix}.
\end{align}
Now, in the Hilbert space $\{|0,0\rangle, |1,0\rangle, |1,\pm1\rangle\}$, the dipole-dipole interaction (DDI), in the second-quantized form, reads

\begin{align}
\hat {\cal H}_{\rm dd}=&\; \frac{g_d}{2}\sqrt{\frac{16\pi}{45}}\int
\frac{d {\bf r}_{1} d {\bf r}_{2}}{|{\mathbf R}|^3}
\left\{Y_{20}(\hat{\mathbf R})\left[\hat{\psi}_{00}^\dag({\mathbf
r}_{1})\hat{\psi}_{11}^\dag({\mathbf r}_{2})\hat{\psi}_{00}({\mathbf
r}_{2})\hat{\psi}_{11}({\mathbf r}_{1}) +\hat{\psi}_{00}^\dag({\mathbf
r}_{1})\hat{\psi}_{1-1}^\dag({\mathbf r}_{2})\hat{\psi}_{00}({\mathbf
r}_{2})\hat{\psi}_{1-1}({\mathbf r}_{1})\right.\right.\nonumber\\
& \qquad\qquad\qquad\qquad\quad\;\left.- 2\hat{\psi}_{00}^\dag({\mathbf r}_{1})\hat{\psi}_{10}^\dag({\mathbf
r}_{2})\hat{\psi}_{00}({\mathbf r}_{2})\hat{\psi}_{1,0}({\mathbf
r}_{1})\right]-Y_{20}(\hat{\mathbf R})\left[\hat{\psi}_{00}^\dag({\mathbf
r}_{1})\hat{\psi}_{00}^\dag({\mathbf r}_{2})\hat{\psi}_{1-1}({\mathbf
r}_{2})\hat{\psi}_{11}({\mathbf r}_{1})\right.\nonumber \\
& \qquad\qquad\qquad\qquad\quad\;\left.\left.+ \hat{\psi}_{00}^\dag({\mathbf r}_{1})\hat{\psi}_{00}^\dag({\mathbf
r}_{2})\hat{\psi}_{10}({\mathbf r}_{2})\hat{\psi}_{10}({\mathbf r}_{1}) +
{\rm h.c.}\right] \right\}\nonumber\\
& -\frac{g_d}{2}\sqrt{\frac{16\pi}{15}}\int \frac{d {\bf r}_{1} d {\bf r}_{2}}{|{\mathbf R}|^3}
\left\{Y_{2-1}(\hat{\mathbf R})\left[\hat{\psi}_{00}^\dag({\mathbf
r}_{1})\hat{\psi}_{00}^\dag({\mathbf r}_{2})\hat{\psi}_{1-1}({\mathbf
r}_{2})\hat{\psi}_{10}({\mathbf r}_{1}) +\hat{\psi}_{00}^\dag({\mathbf
r}_{1})\hat{\psi}_{10}^\dag({\mathbf r}_{2})\hat{\psi}_{00}({\mathbf
r}_{2})\hat{\psi}_{1-1}({\mathbf r}_{1})\right.\right.\nonumber \\
&\qquad\qquad\qquad\qquad\qquad\;\left.\left.
-\hat{\psi}_{00}^\dag({\mathbf r}_{1})\hat{\psi}_{11}^\dag({\mathbf r}_{2})\hat{\psi}_{00}({\mathbf r}_{2})\hat{\psi}_{10}({\mathbf r}_{1}) - \hat{\psi}_{11}^\dag({\mathbf r}_{1})\hat{\psi}_{10}^\dag({\mathbf
r}_{2})\hat{\psi}_{00}({\mathbf r}_{2})\hat{\psi}_{00}({\mathbf r}_{1})\right] +
{\rm h.c.}\right\}\nonumber\\
& -\frac{g_d}{2}\sqrt{\frac{8\pi}{15}}\int\frac{d {\bf r}_{1} d {\bf r}_{2}}{|{\mathbf R}|^3}
\left\{Y_{2-2}(\hat{\mathbf R})\left[\hat{\psi}_{00}^\dag({\mathbf
r}_{1})\hat{\psi}_{00}^\dag({\mathbf r}_{2})\hat{\psi}_{1,-1}({\mathbf r}_{2})\hat{\psi}_{1,-1}({\mathbf r}_{1}) -2\hat{\psi}_{11}^\dag({\mathbf r}_{1})\hat{\psi}_{00}^\dag({\mathbf
r}_{2})\hat{\psi}_{1-1}({\mathbf r}_{2})\hat{\psi}_{00}({\mathbf r}_{1})\right.\right. \nonumber \\
&\qquad\qquad\qquad\qquad\quad\;\;\;
\left.\left.+\hat{\psi}_{11}^\dag({\mathbf r}_{1})\hat{\psi}_{11}^\dag({\mathbf r}_{2})\hat{\psi}_{00}({\mathbf r}_{2})\hat{\psi}_{00}({\mathbf r}_{1})\right] + {\rm h.c.}\right\},\label{smhdd0}
\end{align}
where we have arranged all terms according to the components of the spherical harmonics. From Eq. (\ref{smhdd0}), it is apparent that the DDI conserves the total (rotational + orbital) angular momentum. Next, in the presence of the microwave field, we apply the same unitary transformation, Eq. (\ref{smunitary}), which yields the DDI Hamiltonian in the rotating frame as
\begin{align} 
\hat {\cal H}_{\rm dd}\rightarrow &\;{\cal U}^{\dag}\hat{\cal H}_{\rm dd}{\cal U}\nonumber\\
=&\; \frac{g_d}{2}\sqrt{\frac{16\pi}{45}}\int
\frac{d {\bf r}_{1} d {\bf r}_{2}}{|{\mathbf R}|^3}
\left\{Y_{20}(\hat{\mathbf R})\left[\hat{\psi}_{00}^\dag({\mathbf
r}_{1})\hat{\psi}_{11}^\dag({\mathbf r}_{2})\hat{\psi}_{00}({\mathbf
r}_{2})\hat{\psi}_{11}({\mathbf r}_{1}) +\hat{\psi}_{00}^\dag({\mathbf
r}_{1})\hat{\psi}_{1-1}^\dag({\mathbf r}_{2})\hat{\psi}_{00}({\mathbf
r}_{2})\hat{\psi}_{1-1}({\mathbf r}_{1})\right.\right.\nonumber\\
& \qquad\qquad\qquad\qquad\quad\;\left.- 2\hat{\psi}_{00}^\dag({\mathbf r}_{1})\hat{\psi}_{1,0}^\dag({\mathbf
r}_{2})\hat{\psi}_{00}({\mathbf r}_{2})\hat{\psi}_{10}({\mathbf
r}_{1})\right]-Y_{20}(\hat{\mathbf R})\left[\hat{\psi}_{00}^\dag({\mathbf
r}_{1})\hat{\psi}_{00}^\dag({\mathbf r}_{2})\hat{\psi}_{1-1}({\mathbf
r}_{2})\hat{\psi}_{11}({\mathbf r}_{1})e^{-2i\omega_{\rm mw}t}\right.\nonumber \\
& \qquad\qquad\qquad\qquad\quad\;\left.\left.+ \hat{\psi}_{00}^\dag({\mathbf r}_{1})\hat{\psi}_{00}^\dag({\mathbf
r}_{2})\hat{\psi}_{10}({\mathbf r}_{2})\hat{\psi}_{10}({\mathbf r}_{1})e^{-2i\omega_{\rm mw}t} +
{\rm h.c.}\right] \right\}\nonumber\\
& -\frac{g_d}{2}\sqrt{\frac{16\pi}{15}}\int \frac{d {\bf r}_{1} d {\bf r}_{2}}{|{\mathbf R}|^3}
\left\{Y_{2-1}(\hat{\mathbf R})\left[\hat{\psi}_{00}^\dag({\mathbf
r}_{1})\hat{\psi}_{00}^\dag({\mathbf r}_{2})\hat{\psi}_{1-1}({\mathbf
r}_{2})\hat{\psi}_{10}({\mathbf r}_{1})e^{-2i\omega_{\rm mw}t} +\hat{\psi}_{00}^\dag({\mathbf
r}_{1})\hat{\psi}_{10}^\dag({\mathbf r}_{2})\hat{\psi}_{00}({\mathbf
r}_{2})\hat{\psi}_{1-1}({\mathbf r}_{1})\right.\right.\nonumber \\
&\qquad\qquad\qquad\qquad\qquad\;\left.\left.
-\hat{\psi}_{00}^\dag({\mathbf r}_{1})\hat{\psi}_{11}^\dag({\mathbf r}_{2})\hat{\psi}_{00}({\mathbf r}_{2})\hat{\psi}_{10}({\mathbf r}_{1}) - \hat{\psi}_{11}^\dag({\mathbf r}_{1})\hat{\psi}_{10}^\dag({\mathbf
r}_{2})\hat{\psi}_{00}({\mathbf r}_{2})\hat{\psi}_{00}({\mathbf r}_{1})e^{2i\omega_{\rm mw}t}\right] +
{\rm h.c.}\right\}\nonumber\\
& -\frac{g_d}{2}\sqrt{\frac{8\pi}{15}}\int\frac{d {\bf r}_{1} d {\bf r}_{2}}{|{\mathbf R}|^3}
\left\{Y_{2-2}(\hat{\mathbf R})\left[\hat{\psi}_{00}^\dag({\mathbf
r}_{1})\hat{\psi}_{00}^\dag({\mathbf r}_{2})\hat{\psi}_{1-1}({\mathbf r}_{2})\hat{\psi}_{1-1}({\mathbf r}_{1})e^{-2i\omega_{\rm mw}t} -2\hat{\psi}_{11}^\dag({\mathbf r}_{1})\hat{\psi}_{00}^\dag({\mathbf
r}_{2})\hat{\psi}_{1-1}({\mathbf r}_{2})\hat{\psi}_{00}({\mathbf r}_{1})\right.\right. \nonumber \\
&\qquad\qquad\qquad\qquad\quad\;\;\;
\left.\left.+\hat{\psi}_{11}^\dag({\mathbf r}_{1})\hat{\psi}_{11}^\dag({\mathbf r}_{2})\hat{\psi}_{00}({\mathbf r}_{2})\hat{\psi}_{00}({\mathbf r}_{1})e^{2i\omega_{\rm mw}t}\right] + {\rm h.c.}\right\}.\label{smhdd1}
\end{align}
As estimated in the main text, the rotational splitting $2B_{v}$ is much larger than the DDI energy for a typical gas density. Consequently, the spin dynamics induced by the DDI is much slower than the Rabi oscillations induced by the microwave field. We may therefore use an effective DDI which is time-averaged over a period of $2\pi/\omega_{\rm mw}$,
\begin{align} 
\hat {\cal H}_{\rm dd}\simeq&\; \frac{g_d}{2}\sqrt{\frac{16\pi}{45}}\int
\frac{d {\bf r}_{1} d {\bf r}_{2}}{|{\mathbf R}|^3}
\left\{Y_{20}(\hat{\mathbf R})\left[\hat{\psi}_{00}^\dag({\mathbf
r}_{1})\hat{\psi}_{11}^\dag({\mathbf r}_{2})\hat{\psi}_{00}({\mathbf
r}_{2})\hat{\psi}_{11}({\mathbf r}_{1}) +\hat{\psi}_{00}^\dag({\mathbf
r}_{1})\hat{\psi}_{1-1}^\dag({\mathbf r}_{2})\hat{\psi}_{00}({\mathbf
r}_{2})\hat{\psi}_{1-1}({\mathbf r}_{1})\right.\right.\nonumber\\
& \qquad\qquad\qquad\qquad\quad\;\left.\left.- 2\hat{\psi}_{00}^\dag({\mathbf r}_{1})\hat{\psi}_{1,0}^\dag({\mathbf
r}_{2})\hat{\psi}_{00}({\mathbf r}_{2})\hat{\psi}_{10}({\mathbf
r}_{1})\right] \right\}\nonumber\\
& -\frac{g_d}{2}\sqrt{\frac{16\pi}{15}}\int \frac{d {\bf r}_{1} d {\bf r}_{2}}{|{\mathbf R}|^3}
\left\{Y_{2-1}(\hat{\mathbf R})\left[\hat{\psi}_{00}^\dag({\mathbf
r}_{1})\hat{\psi}_{10}^\dag({\mathbf r}_{2})\hat{\psi}_{00}({\mathbf
r}_{2})\hat{\psi}_{1-1}({\mathbf r}_{1})
-\hat{\psi}_{00}^\dag({\mathbf r}_{1})\hat{\psi}_{11}^\dag({\mathbf r}_{2})\hat{\psi}_{00}({\mathbf r}_{2})\hat{\psi}_{10}({\mathbf r}_{1})\right] +
{\rm h.c.}\right\}\nonumber\\
& -\frac{g_d}{2}\sqrt{\frac{8\pi}{15}}\int\frac{d {\bf r}_{1} d {\bf r}_{2}}{|{\mathbf R}|^3}
\left[-2Y_{2-2}(\hat{\mathbf R})\hat{\psi}_{11}^\dag({\mathbf r}_{1})\hat{\psi}_{00}^\dag({\mathbf
r}_{2})\hat{\psi}_{1-1}({\mathbf r}_{2})\hat{\psi}_{00}({\mathbf r}_{1})+ {\rm h.c.}\right].\label{smhdd2}
\end{align}
The adiabatic elimination of the $|0,0\rangle$ level from the interaction Hamiltonian (\ref{smhdd2}) can be achieved by simply performing the substitution ${\hat \psi}_{00} = -{\Omega{\hat \psi}_{11}}/{\Delta}$, which leads to 
\begin{align} 
\hat {\cal H}_{\rm dd}\simeq&\; \frac{\kappa g_d}{2}\sqrt{\frac{16\pi}{45}}\int
\frac{d {\bf r}_{1} d {\bf r}_{2}}{|{\mathbf R}|^3}
\left\{Y_{20}(\hat{\mathbf R})\left[\hat{\psi}_{11}^\dag({\mathbf
r}_{1})\hat{\psi}_{11}^\dag({\mathbf r}_{2})\hat{\psi}_{11}({\mathbf
r}_{2})\hat{\psi}_{11}({\mathbf r}_{1}) +\hat{\psi}_{11}^\dag({\mathbf
r}_{1})\hat{\psi}_{1-1}^\dag({\mathbf r}_{2})\hat{\psi}_{11}({\mathbf
r}_{2})\hat{\psi}_{1-1}({\mathbf r}_{1})\right.\right.\nonumber\\
& \qquad\qquad\qquad\qquad\quad\;\left.\left.- 2\hat{\psi}_{11}^\dag({\mathbf r}_{1})\hat{\psi}_{1,0}^\dag({\mathbf
r}_{2})\hat{\psi}_{11}({\mathbf r}_{2})\hat{\psi}_{10}({\mathbf
r}_{1})\right] \right\}\nonumber\\
& -\frac{\kappa g_d}{2}\sqrt{\frac{16\pi}{15}}\int \frac{d {\bf r}_{1} d {\bf r}_{2}}{|{\mathbf R}|^3}
\left\{Y_{2-1}(\hat{\mathbf R})\left[\hat{\psi}_{11}^\dag({\mathbf
r}_{1})\hat{\psi}_{10}^\dag({\mathbf r}_{2})\hat{\psi}_{11}({\mathbf
r}_{2})\hat{\psi}_{1-1}({\mathbf r}_{1})
-\hat{\psi}_{11}^\dag({\mathbf r}_{1})\hat{\psi}_{11}^\dag({\mathbf r}_{2})\hat{\psi}_{11}({\mathbf r}_{2})\hat{\psi}_{10}({\mathbf r}_{1})\right] +
{\rm h.c.}\right\}\nonumber\\
& -\frac{\kappa g_d}{2}\sqrt{\frac{8\pi}{15}}\int\frac{d {\bf r}_{1} d {\bf r}_{2}}{|{\mathbf R}|^3}
\left[-2Y_{2-2}(\hat{\mathbf R})\hat{\psi}_{11}^\dag({\mathbf r}_{1})\hat{\psi}_{11}^\dag({\mathbf
r}_{2})\hat{\psi}_{1-1}({\mathbf r}_{2})\hat{\psi}_{11}({\mathbf r}_{1})+ {\rm h.c.}\right].\label{smhdd3}
\end{align}
As can be seen, the elimination of $|0,0\rangle$ level gives rise to the factor $\kappa$ to the DDI strength which can be used as a control knob for the DDI. For the parameter regime considered in this work, the DDI energy $\kappa g_{d}n$ is much smaller than the level splitting between $|1,0\rangle$ and $|1,\pm1\rangle$. Therefore, with the assumption that all molecules are prepared in the $|1,1\rangle$ state, the $|1,0\rangle$ level is essentially unoccupied during the time scale considered here. As a result, we may simply drop all terms containing $\hat\psi_{10}$ in $\hat{\cal H}_{\rm dd}$, which eventually leads to the effective DDI Hamiltonian in the main text.

\end{widetext}

\end{document}